\documentclass[superscriptaddress, prd, aps,amsmath,amssymb,showpacs,showkeys, onecolumn]{revtex4-2}
\usepackage[dvips]{graphicx,color}
\usepackage{subfig}
\usepackage{times}
\usepackage{xcolor}
\usepackage[%
  colorlinks=true,
  urlcolor=blue,
  linkcolor=red,
  citecolor=blue
]{hyperref}
\usepackage{orcidlink}

\begin{document}
\title{Thermodynamics of deformed AdS-Schwarzschild black holes in the presence of Thermal fluctuations}

\author{Dhruba Jyoti Gogoi\orcidlink{0000-0002-4776-8506}}
\email[Email: ]{moloydhruba@yahoo.in}

\affiliation{Department of Physics, Moran College, Moranhat, Charaideo 785670, Assam, India}
\affiliation{Research Center of Astrophysics and Cosmology, Khazar University, 41 Mehseti Street, AZ1096 Baku, Azerbaijan}
\affiliation{Theoretical Physics Division, Centre for Atmospheric Studies, Dibrugarh University, Dibrugarh
786004, Assam, India}

\author{Poppy Hazarika\orcidlink{0000-0002-1179-1528}}
\email[Email: ]{ poppyhazarika1@gmail.com}
\affiliation{Department of Physics, Duliajan College, Duliajan 786602, Assam, India}

\author{Jyatsnasree Bora \orcidlink{0000-0001-9751-5614}}
\email[Email: ]{jyatnasree.borah@gmail.com}

\affiliation{Department of Physics, D.H.S.K. College, Dibrugarh, 786001, Assam, India}

\author{Ranjan Changmai}
\email[Email: ]{rchangmai12@gmail.com}

\affiliation{Dibru College,
Dibrugarh 786003, Assam, India}

\begin{abstract}
This paper examines the thermodynamic properties and stability of deformed AdS-Schwarzschild black holes, focusing on the effects of deformation ($\alpha$) and thermal correction parameters ($\beta_1$, $\beta_2$) on phase transitions and heat capacity. The results show that higher $\alpha$ values raise the Hawking-Page critical temperature, enhancing thermal stability. Thermal corrections significantly affect smaller black holes but minimally impact larger ones, leaving second-order phase transitions unchanged. Heat capacity analysis identifies stability regions, with sign changes marking instability. These findings highlight the role of deformation and thermal corrections in black hole stability, offering insights for extending our understanding of black hole thermodynamics.

\end{abstract}


\maketitle
\section{Introduction}\label{sec01}
 
 The study of black holes has long intrigued scientists, not only from a gravitational perspective but also from a thermodynamic one. Black hole thermodynamics, a fascinating intersection of classical thermodynamics, general relativity, quantum mechanics, and statistical mechanics, has emerged as a fundamental area of research in theoretical physics. This field seeks to extend the principles of thermodynamics to the realm of black holes, shedding light on the profound implications these objects have on our understanding of the universe. Since { Bekenstein's} seminal work in 1973, in which he first proposed that black holes have an entropy proportional to their event horizon area \cite{Bekenstein1,Bekenstein2}, black holes have been recognized as thermodynamic objects that obey the laws of thermodynamics.

Building on Bekenstein's insights, S. Hawking in 1974 introduced the concept of Hawking radiation, demonstrating that black holes emit radiation with a spectrum resembling that of a black body at a specific temperature. This temperature is proportional to the surface gravity of the black hole \cite{Belgiorno}, and the radiation was named after Hawking. Hawking's findings revealed that black holes are not entirely isolated objects; instead, they lose mass and energy through radiation, challenging the classical idea that nothing can escape a black hole. These discoveries led to the development of the four laws of black hole thermodynamics, which closely parallel the classical laws of thermodynamics \cite{Bardeen}. As a result, black hole thermodynamics has gained significant momentum in research, incorporating new parameters such as heat capacity, pressure, and chemical potential into the study of black holes \cite{Wald, Carlip}.

One of the key cornerstones of black hole thermodynamics is the Bekenstein-Hawking area law, which posits that the entropy of a black hole is proportional to the area of its event horizon and is given by $ S = A / 4 $ \cite{Bekenstein1, Bardeen}. However, this law primarily applies to large black holes in equilibrium. For small-sized black holes, where Hawking radiation significantly influences their dynamics, quantum fluctuations around the equilibrium state necessitate corrections to the entropy. These quantum corrections, derived from various approaches, generally introduce a logarithmic term as a first-order correction to the black hole's entropy \cite{Mandal}. In 2001, D. Birmingham and S. Sen applied Rademacher expansion to derive logarithmic corrections to black hole entropy within a conformal field theory framework, providing deeper insight into these modifications \cite{Birmingham}. { In another recent work, the authors investigated the tunneling radiation as well as thermal fluctuations of a gauge super gravity like black hole \cite{rali}.}

Such corrections have profound implications for the thermodynamics of various black hole solutions. Elaborate studies have explored how these corrections affect rotating and charged black holes in different spacetimes, such as BTZ black holes, massive black holes in AdS spaces, and dilatonic black holes \cite{Islam, Upadhyay, Upadhyay1, Upadhyay2, Pourhassan}. Furthermore, the effects of corrected entropy have been extended to study specific black hole solutions, such as the Schwarzschild-Beltrami-de Sitter black hole and the Goedel black hole \cite{Upadhyay5, Islam1, Jing}. These works highlight that small-sized black holes are more susceptible to quantum corrections, with first-order corrections often destabilizing them, while second-order corrections tend to enhance stability \cite{Ndongmo, Gogoi}.

The thermodynamic properties of black holes are also strongly influenced by the presence of external fields and the cosmological constant. Electromagnetic fields and a positive or negative cosmological constant can affect black hole thermodynamics, leading to modified expressions for entropy and temperature \cite{Kubiznak:2012wp,Hendi:2012um,Teitelboim:1985dp, Arenas-Henriquez:2022ntz}. The study of phase transitions in black holes, akin to those in conventional thermodynamic systems, has become a key focus. For instance, Hawking and Page first explored the phase transitions of AdS-Schwarzschild black holes, revealing a critical temperature beyond which a phase transition occurs \cite{Hawking}. Such transitions are now a common feature in studies of black holes within modified gravity theories, where a wide variety of phase transitions have been discovered across different black hole types \cite{Wei:2012ui, Altamirano:2014tva}.

Incorporating insights from holography, black hole thermodynamics continues to evolve. Holography, particularly in the context of the AdS/CFT correspondence, has opened new avenues for understanding the thermodynamic behaviour of black holes. For example, rotating Kerr-AdS black holes have been analyzed for their phase transitions and thermodynamic properties \cite{Gong:2023ywu}. The study of microstructures and stability in black holes within the AdS/CFT framework has further enhanced our understanding of these complex objects \cite{Sokoliuk:2023pby, Singh:2023hit}.

Recent developments in the field have also explored the impact of quantum fluctuations on black holes in GR and different modified theories of gravity \cite{Anacleto:2015mma, Anacleto:2020lel}. These studies often lead to deformed expressions for black hole entropy and temperature, which could yield potential observational effects, providing a deeper understanding of quantum gravity \cite{Ashtekar:2004eh, Dvali:2011aa, Ali:2009zq, Yang:2023nnk, Wang:2025fmz, Battista:2023iyu}. The thermodynamics of black holes, thus, remains a crucial tool in unravelling the mysteries of quantum gravity and understanding the fundamental nature of spacetime \cite{Khosravipoor_2023, hawp, More:2004hv, PF, SPR, Mandal:2023ahb, Sahabandu:2005ma, Cai:2003kt, Bekenstein3, Upadhyay3, Pourhassan1, Upadhyay4, Upadhyay6, Pourdarvish, Pourhassan2, Govindarajan, Chamblin, Pourhassan4, Pourhassan5, Pourhassan6, Mancilla:2024spp, Karch:2015rpa, Anacleto:2020zfh, Ali:2023dku, Ali:2023jbj, Akhtar:2022jjk,Ditta:2024pxl,Feng:2024jeq, Sekhmani:2024frn, Ditta:2024tdo, Sekhmani:2025fpw, Sekhmani:2024kfj}.

The motivation behind our work is to study the effect of thermal fluctuations on thermodynamic properties of the deformed AdS black hole system \cite{Khosravipoor_2023} and thereby analyse the stability and behaviour of these systems under such perturbations.
In our work, first the basic thermodynamic variables of the AdS black hole systems are constructed. Then, the second-order corrections to entropy of the black holes in the presence of thermal fluctuations are made. With the corrected entropy, the thermodynamic potentials like enthalpy, Helmholtz free energy, thermodynamic volume, specific heat, internal energy, and Gibbs’ free energy of the deformed AdS black hole systems are calculated. The introduction of deformation to the AdS black hole system influences the black hole's phase transition characteristics and hence its thermal stability. Thus such studies will enrich our knowledge on the deformed black hole thermodynamics in the context of AdS spacetimes.  
  
\section{Deformed AdS-Schwarzschild black hole}

To construct the deformed AdS black hole solution, we begin with the following four-dimensional action \cite{Khosravipoor_2023}:

\begin{equation}\label{main action}
    A=\int{d^4 x\,
    \sqrt{-g}\,\left(\frac{R-2\Lambda}{2\kappa}+\mathcal{L}_m+\mathcal{L}_{\rm X}\right)},
\end{equation}

where $ g $ represents the determinant of the metric, $ R $ is the Ricci scalar, $ \Lambda $ is the cosmological constant, and $ \kappa=8\pi G/c^4 $. Here, $ \mathcal{L}_m $ refers to the matter Lagrangian, and $ \mathcal{L}_{\rm X} $ represents the Lagrangian corresponding to theories beyond General Relativity (GR) or additional fields such as scalar, vector, or tensor fields. In the absence of the extra Lagrangian term $ \mathcal{L}_{\rm X} $, the field equation reduces to

\begin{equation}\label{enesteineq}
    G_{\mu \nu} + \Lambda g_{\mu\nu} = \kappa\, T^{(\rm tot)}_{\mu\nu},
\end{equation}

which leads to the standard Schwarzschild-AdS metric solution in the absence of matter,

\begin{equation}\label{Ads}
    e^{\zeta(r)}|_{T^{(\rm m)}_{\mu\nu}=0}=e^{-\lambda(r)}|_{T^{(\rm m)}_{\mu\nu}=0}=1-\frac{2M}{r}+\frac{r^2}{l^2}.
\end{equation}

In the above, $ G_{\mu\nu} $ is the Einstein tensor, and the total stress-energy tensor is represented as the sum of contributions from matter and additional fields beyond GR:

\begin{equation}\label{ems}
    T^{(\rm tot)}_{\mu\nu} = T^{(\rm m)}_{\mu\nu} + T^{(\rm X)}_{\mu\nu}.
\end{equation}

Now we follow Ref. \cite{Khosravipoor_2023} to obtain a deformed AdS black hole solution. To incorporate the deformation, an energy density function proportional to $ 1/r^4 $ is introduced as given below \cite{Khosravipoor_2023}:

\begin{equation}\label{Efunction}
    \mathcal{E}(r)=\frac{\alpha }{\kappa  (\beta +r)^4},
\end{equation}

where $ \beta $ is a constant parameter controlling the behaviour of the energy density at $ r=0 $, and $ \alpha $ is the deformation parameter. This function ensures a rapid asymptotic decay and avoids central singularity.

By substituting this energy density into the field equations, and solving for the metric function one can obtain:

\begin{equation}\label{metric function}
  F(r)=1-\frac{2M}{r}+\frac{r^2}{l^2} +\alpha\,\frac{\beta^2+3r^2+3\beta r}{3r(\beta+r)^3}.
\end{equation}
{ In the above metric function, $l^2 =-\frac{3}{\Lambda} $.}
Thus, the final line element for the deformed AdS-Schwarzschild black hole reads:

\begin{equation}\label{final line element}
    ds^2 = - F(r) dt^2 + \frac{1}{F(r)} dr^2 + r^2(d\theta^2 + \sin^2{\theta}\, d\phi^2).
\end{equation}

This metric describes a black hole solution with a deformation parameter $ \alpha $ that preserves the asymptotic AdS behaviour and introduces modifications consistent with additional gravitational fields. In the following sections, we will explore the horizon structure and thermodynamic properties of this deformed AdS black hole.
{ One may note that this deformed black hole is not a regular black hole or is not free from singularity. One may clearly observe that the terms associated with the black hole mass $M$ and deformation parameter carry singularity issue.}

\section{Hawking Temperature and Basic Thermodynamic Parameters}\label{sec3}

This section deals with the basic thermodynamic variables of the deformed AdS black holes. At first, we start with the volume of the black hole which can be obtained from the event horizon radius of the black hole by solving the Eq. $f(r)=0$ at the event horizon radius $r_+$. The volume can be given by:

\begin{equation}
    V = \frac{4 \pi  r_+^3}{3}.
\end{equation}

Now, one can obtain the Hawking temperature $T_H$ of the black hole with the help of the lapse function defined by Eq.(\ref{metric function}) as follows,
\begin{equation}\label{Th}
T_{H}=\frac{f^\prime(r)}{4\pi}\Bigg|_{r\ = \ r_{+}}=\frac{\beta ^4-\alpha  r_+^2+4 \beta ^3 r_++6 \beta ^2 r_+^2-\Lambda  r_+^2 \left(\beta +r_+\right){}^4+4 \beta  r_+^3+r_+^4}{4 \pi  r_+ \left(\beta +r_+\right){}^4},
\end{equation}
where the prime ``$\prime$" stands for differentiation w.r.t. $r$. In the GR limit, the above expression for Hawking temperature reduces to the Hawking temperature of a Schwarzschild black hole. 
\begin{equation}
\lim_{(\Lambda, \alpha, \beta)\to(0,0, 0)} T_{H} = \frac{1}{8\pi M}.
\end{equation}

\begin{figure*}[t!]
      	\centering{
       \includegraphics[scale=0.65]{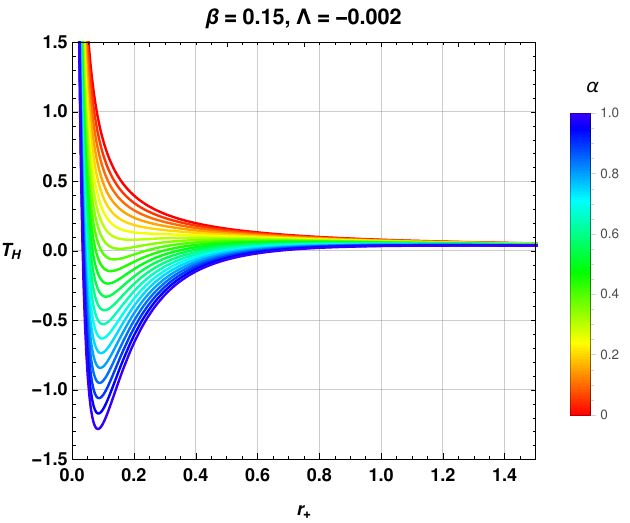} \hspace{5mm}
       \includegraphics[scale=0.65]{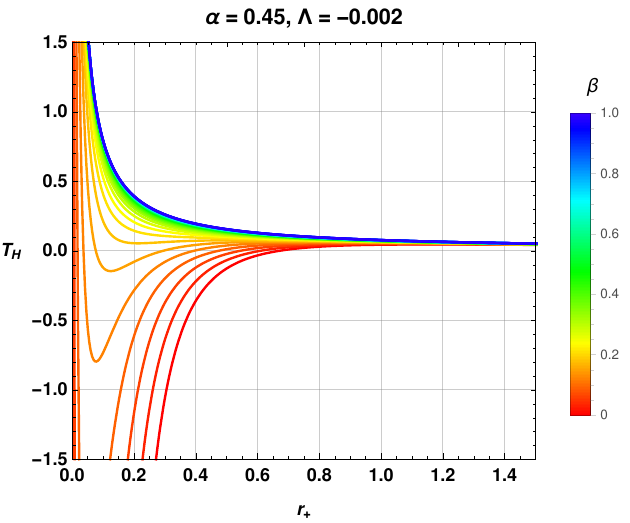}
       
       }
      	\caption{Variation of the Hawking temperature with the black hole horizon radius $r_+$.}
      	\label{figT01}
      \end{figure*}

\begin{figure*}[t!]
      	\centering{
      	\includegraphics[scale=0.65]{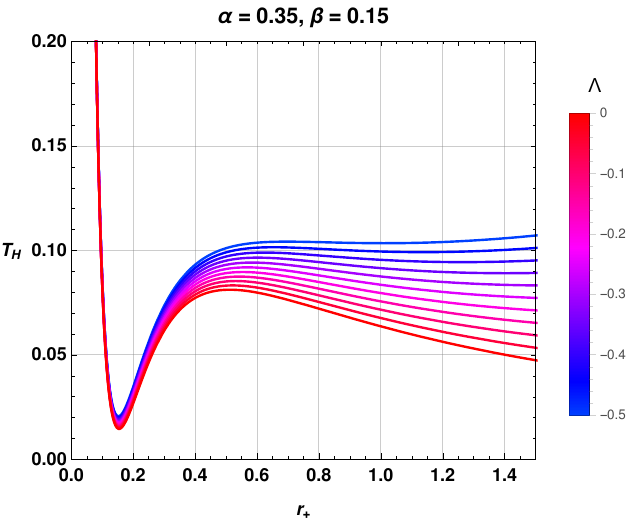}
       
       }
      	\caption{Variation of the Hawking temperature with the black hole horizon radius $r_+$.}
      	\label{figT02}
      \end{figure*}

The dependency of Hawking temperature on the model parameters is depicted in Fig.s \ref{figT01} and \ref{figT02}. In the first panel of Fig. \ref{figT01}, we have shown the impacts of the deformation parameter $\alpha$ on the Hawking radiation $T_H$. One can see that with an increase in the value of $\alpha$, the Hawking radiation becomes negative for smaller black holes. However, as $r_+ \rightarrow 0$, temperature increases drastically. Another observation is that the impact of the deformation parameter decreases significantly for larger black holes. In the second panel of Fig. \ref{figT01}, we have shown the impact of the control parameter $\beta$ on the Hawking temperature of the black hole. It is seen that the impact of $\beta$ on the Hawking temperature is opposite in comparison to that of $\alpha$. A higher value of $\beta$ makes the Hawking temperature behaviour similar to that of a Schwarzschild black hole. Finally, we have shown the impacts of cosmological constant $\Lambda$, which is associated with the pressure $P$ of the black hole, on the Hawking temperature in Fig \ref{figT02}.

The thermodynamic properties of deformed AdS black holes can be analyzed through various key quantities, starting with the Bekenstein entropy. The Bekenstein entropy, which is a fundamental concept in black hole thermodynamics, is calculated using the laws of black hole thermodynamics. For a deformed AdS black hole, the uncorrected entropy is given by the expression:
\begin{equation}\label{eq31}
    S _0= \pi r^2_+  .
\end{equation}
Moving on to the enthalpy energy of the system, it is determined through the integration of temperature $T_H$ with respect to the entropy $S_0$ i.e.,
\begin{equation}\label{eq32}
     H = \int T_HdS_0.
\end{equation}
For this black hole system, the enthalpy is found to be
\begin{equation}\label{eq33}
    H = \frac{1}{6} \left(3 \beta +\frac{\alpha  \left(\beta ^2+3 \beta  r_++3 r_+^2\right)}{\left(\beta +r_+\right){}^3}-\Lambda  \left(\beta ^3+r_+^3\right)+3 r_+\right).
\end{equation}
One can see that the enthalpy is a function of both the model parameters $\alpha,$ the deformation parameter and $\beta,$ the control parameter. Now, the expression for pressure $P$ of the black hole can be written as:
\begin{equation}\label{eq34}
P=-\frac{\Lambda}{8\pi},
\end{equation}
which highlights the direct dependence of pressure on the cosmological constant $\Lambda$.
With the expressions for enthalpy and pressure in hand, one can derive other important thermodynamic properties such as the internal energy $U$, Helmholtz free energy $F$, and specific heat $C_0$ for the black hole system. The internal energy, in particular, is found using the familiar thermodynamic identity $U=H-PV$. For this black hole system, it is:
\begin{equation}
    U = \frac{1}{6} \left(3 \beta +\frac{\alpha  r_+^2}{\left(\beta +r_+\right){}^3}+r_+ \left(\frac{\alpha }{\left(\beta +r_+\right){}^2}+3\right)+\frac{\alpha }{\beta +r_+}\right).
\end{equation}
In the subsequent sections, the focus will shift to examining the effects of small stable fluctuations near equilibrium on the thermodynamic properties of the deformed AdS black hole system. This analysis will provide deeper insights into the stability and behaviour of the system under perturbations, further enriching our understanding of the deformed black hole thermodynamics in the context of AdS spacetimes. 

\section{Thermodynamic fluctuations: the second-order corrections to entropy}\label{sec4}

In this section, we examine the influence of thermal fluctuations on the entropy of black holes. The groundbreaking work by Hawking and Page \cite{hawp} established that black holes in asymptotically curved spacetime can be effectively modelled using a canonical ensemble framework. Building on this approach, we analyze the system of deformed AdS black holes as a canonical ensemble comprising $N$ particles, each with an energy spectrum $E_{n}$. The statistical behaviour of the system is encapsulated by the partition function, which is expressed as:

\begin{equation}
    Z = \int_0^\infty  dE  \rho (E) e^{-\bar{\beta}_{\kappa} E},
\end{equation}
where $\bar{\beta}_{\kappa}$ represents the inverse temperature (in units of the Boltzmann constant), and $\rho (E)$ denotes the canonical density of states corresponding to the average energy $E$. By utilizing the partition function $Z$ and applying Laplace inversion, the density of states can be determined as:

\begin{equation} \label{eq17}
    \rho (E) = \frac{1}{2 \pi i} \int^{\bar{\beta}_{0\kappa}+
i\infty}_{\bar{\beta}_{0\kappa} - i\infty} d \bar{\beta}_{\kappa}  e^{S(\bar{\beta}_{\kappa})},
\end{equation}

where the entropy is given by
\begin{equation}
    S = \bar{\beta}_{\kappa} E + \log Z.
\end{equation}

The entropy around the equilibrium temperature $\bar{\beta}_{0\kappa}$ is obtained by eliminating all thermal fluctuations. However, in the presence of thermal fluctuations, the corrected entropy can be expressed by a Taylor expansion around $\bar{\beta}_{0\kappa}$ as:

\begin{equation} \label{eq19}
    S = S_0 + \frac{1}{2}(\bar{\beta}_{\kappa} - \bar{\beta}_{0\kappa})^2 \left(\frac{\partial^2 S(\bar{\beta}_{\kappa})}{\partial \bar{\beta}_{\kappa}^2 }\right)_{\bar{\beta}_{\kappa} = \bar{\beta}_{0\kappa}} + \frac{1}{6}(\bar{\beta}_{\kappa} - \bar{\beta}_{0\kappa})^3 \left(\frac{\partial^3 S(\bar{\beta}_{\kappa})}{\partial \bar{\beta}_{\kappa}^3 }\right)_{\bar{\beta}_{\kappa} = \bar{\beta}_{0\kappa}} + \cdots,
\end{equation}

where the dots denote higher-order corrections.

We should note that the first derivative of the entropy for $\bar{\beta}_{\kappa}$ vanishes at the equilibrium temperature. Consequently, the density of states can be { obtained by using Eq. \eqref{eq17} and \eqref{eq19} } as \cite{More:2004hv}:
\begin{equation}
    \rho (E) = \frac{e^{S_0}}{2 \pi i} \int^{\bar{\beta}_{0\kappa} + i\infty}_{\bar{\beta}_{0\kappa} - i\infty} d \bar{\beta}_{\kappa} \, \exp \left( \frac{(\bar{\beta}_{\kappa} - \bar{\beta}_{0\kappa})^2}{2} \left(\frac{\partial^2 S(\bar{\beta}_{\kappa})}{\partial \bar{\beta}_{\kappa}^2 }\right)_{\bar{\beta}_{\kappa} = \bar{\beta}_{0\kappa}} + \frac{(\bar{\beta}_{\kappa} - \bar{\beta}_{0\kappa})^3}{6} \left(\frac{\partial^3 S(\bar{\beta}_{\kappa})}{\partial \bar{\beta}_{\kappa}^3 }\right)_{\bar{\beta}_{\kappa} = \bar{\beta}_{0\kappa}} + \cdots \right).
\end{equation}

Following the approach in Ref. \cite{More:2004hv}, we obtain:
\begin{equation}
    S = S_0 - \frac{1}{2} \log{S_{0}T_{H}^{2}} + \frac{f(m,n)}{S_{0}} + \cdots,
\end{equation}

where $f(m,n)$ is considered a constant. A more general expression for the corrected entropy is given by \cite{PF}:
\begin{equation}
    S = S_0 - \frac{\beta_1}{2} \log(S_0 T_{H}^2) + \frac{\beta_2}{S_0} + \cdots,
\end{equation}

where the parameters $\beta_1$ and $\beta_2$ are introduced to track the first-order and second-order corrected terms. When $\beta_1 \rightarrow 0$ and $\beta_2 \rightarrow 0$, the original results are recovered, and $\beta_1 = 1$ and $\beta_2 = 0$ yield the usual corrections \cite{More:2004hv,SPR}. Therefore, the first-order correction is logarithmic, while the second-order correction is proportional to the inverse of the original entropy $S_0$. These corrections can be considered quantum corrections to the black hole. For large black holes, these corrections can be neglected, however, as the black hole decreases in size due to Hawking radiation, the quantum fluctuations in the black hole's geometry increase. { Moreover, in the case of a microscopic black hole with quantum correction, these thermal fluctuation parameters may have significant contributions.} Thus, thermal fluctuations significantly modify the thermodynamics of black holes \cite{Mandal:2023ahb}, becoming more important as the black holes reduce in size.

Up to the second-order correction, the explicit form of the entropy for this black hole is given by
\begin{equation}
    S_c = \pi  r_+^2+ \beta _1 \log (16 \pi )-\beta _1 \log \left(\frac{\left(\beta ^4-\alpha  r_+^2+4 \beta ^3 r_++6 \beta ^2 r_+^2-\Lambda  r_+^2 \left(\beta +r_+\right){}^4+4 \beta  r_+^3+r_+^4\right){}^2}{\left(\beta +r_+\right){}^8}\right)+\frac{\beta _2}{\pi  r_+^2}.
\end{equation}
{ In the above expression of corrected entropy $S_c$, once we set $\beta_1 = \beta_2 = 0$, we obtain the uncorrected entropy $S_0$ in absence of thermal fluctuations. So, the above equation is of the general form: $S_c = S_0 + \text{(correction\; terms\; from\; thermal\; fluctuations)}$.}

\begin{figure*}[t!]
      	\centering{
       \includegraphics[scale=0.65]{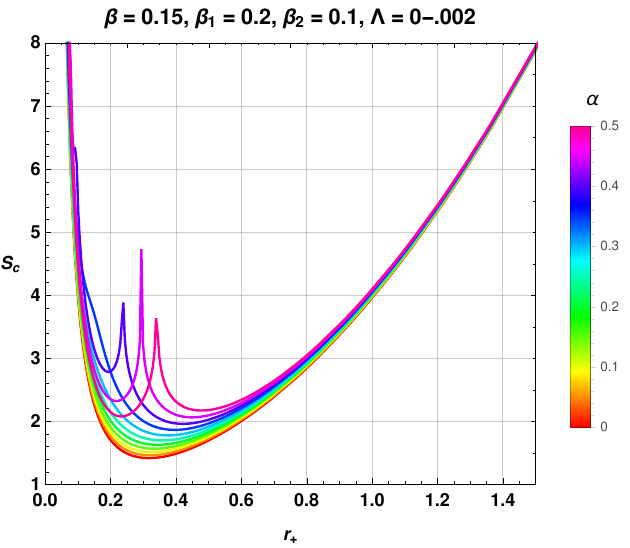} \hspace{5mm}
       \includegraphics[scale=0.65]{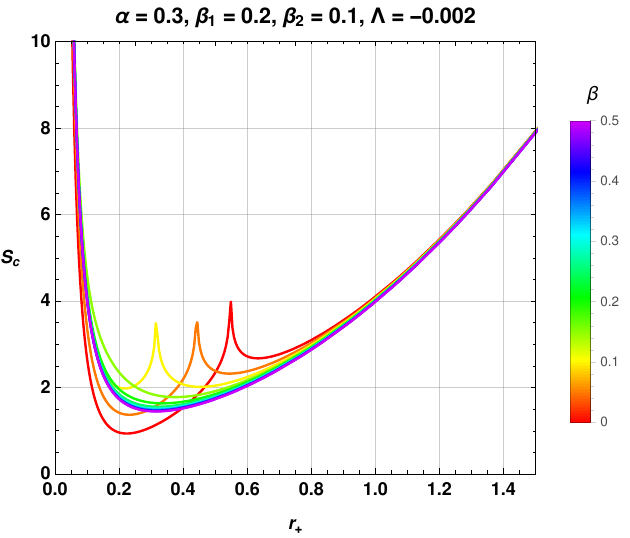}
       
       }
      	\caption{Corrected entropy vs. the black hole horizon radius $r_+$.}
      	\label{figSc01}
      \end{figure*}

\begin{figure*}[t!]
      	\centering{
       \includegraphics[scale=0.65]{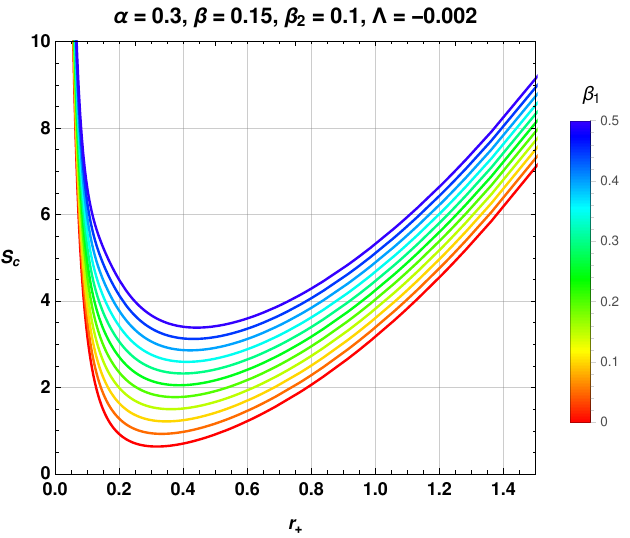} \hspace{5mm}
       \includegraphics[scale=0.65]{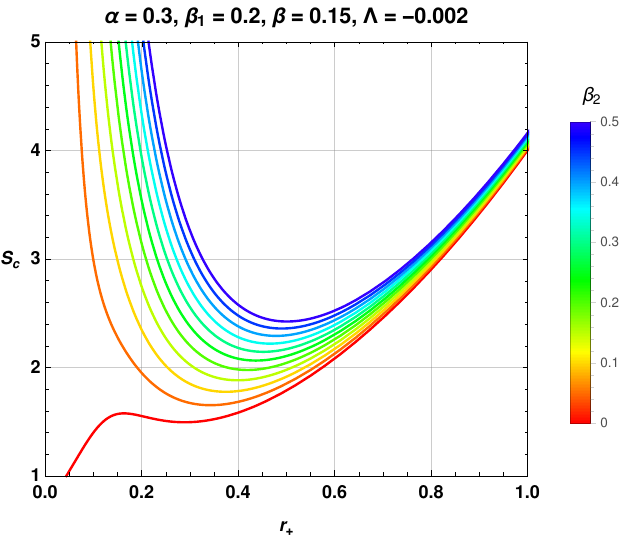}
       
       }
      	\caption{Corrected entropy vs. the black hole horizon radius $r_+$.}
      	\label{figSc02}
      \end{figure*}

The variation of the corrected entropy of the black hole is shown in Fig.s \ref{figSc01} and \ref{figSc02}. Here, we have demonstrated the variations concerning different parameters of this study. One may note that with an increase in the value of the parameter $\alpha$, we observe an increase in the entropy of the black hole with a smaller event horizon. The spikes, which are present due to the thermal fluctuations, shift towards larger $r_+$ with an increase in the value of $\alpha$. On the other hand, the spikes move towards smaller values of $r_+$ with an increase in the parameter $\beta$. For larger values of $\beta$, we do not observe spikes in the entropy curve. From Fig.  \ref{figSc02}, it is clear that the presence of thermal fluctuations increases the entropy of the black hole. However, the impacts of both the fluctuation parameters are different. { Moreover, the spikes, as one can see, are significant for higher values of the parameter $\alpha$ and lower values of the parameter $\beta$. The corrected entropy $S_c$ of a black hole incorporates a second-order correction term $\frac{\beta_2}{\pi r_+^2}$, which gives rise to a weak singularity as the horizon radius $r_+$ nears zero. This divergence highlights the limitations of the semiclassical approximation, suggesting that quantum gravity effects become significant for very small black holes. Physically, this behaviour points to thermodynamic instability—such as unbounded entropy—and hints at a minimal length scale, possibly the Planck scale, beyond which classical spacetime descriptions break down.}

In the next section, we will analyze the thermodynamic quantities of the black hole with the corrected entropy.

\section{Impact of Thermal fluctuations: the Second order corrected thermodynamic potentials}\label{sec5}

This section aims to evaluate various thermodynamic variables for the deformed AdS black hole in the presence of thermodynamic fluctuations. As mentioned earlier, we shall consider the second-order entropy corrections here. Using the second-order corrected entropy, { it is possible to derive corrected enthalpy energy $H_c$ as given by}
\begin{equation}\label{29}
H_{c}=\int T_{H}dS_{c}. 
\end{equation}
Using the expressions of $T_H$ and $S_c$ in the above equation, we can have the corrected enthalpy
\begin{multline}
    H_c = (2 \pi ^2)^{-1} \Big[\frac{1}{3} \Big \lbrace\frac{12 \alpha  \beta _2 \log \left(\beta +r_+\right)}{\beta ^5}+\pi ^2 \left(\frac{\alpha  \left(\beta ^2+3 \beta  r_++3 r_+^2\right)}{\left(\beta +r_+\right){}^3}-\Lambda  r_+^3+3 r_+\right)+\pi  \beta _1 \left(\frac{\alpha  \left(2 r_+-\beta \right)}{\left(\beta +r_+\right){}^4}+6 \Lambda  r_+\right) \\ +\beta _2 \left(-\frac{3 \left(\alpha +\beta ^4 \Lambda \right)}{\beta ^4 r_+}-\frac{\alpha  \left(13 \beta ^2+21 \beta  r_++9 r_+^2\right)}{\beta ^4 \left(\beta +r_+\right){}^3}+\frac{1}{r_+^3}\right)\Big\rbrace-\frac{4 \alpha  \beta _2 \log \left(r_+\right)}{\beta ^5} \Big].
    \end{multline}

\begin{figure*}[t!]
      	\centering{
       \includegraphics[scale=0.85]{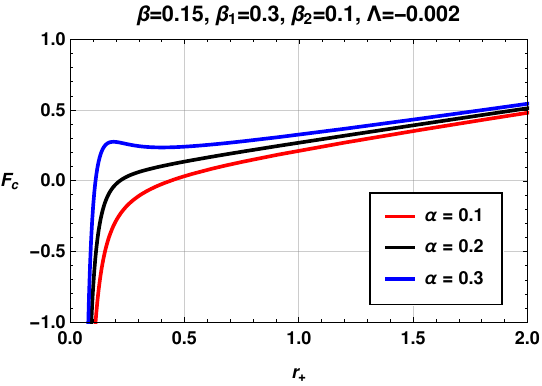} \hspace{5mm}
       \includegraphics[scale=0.85]{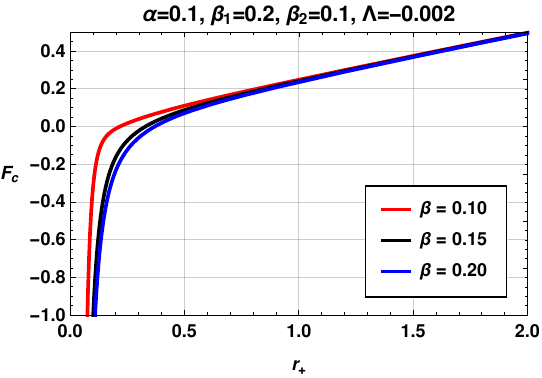}
       
       }
      	\caption{Variation of the corrected Helmholtz energy of the black hole with the event horizon radius $r_+$ for different values of the deformation parameter $\alpha$ and control parameter $\beta$.}
      	\label{figFc01}
      \end{figure*}

\begin{figure*}[t!]
      	\centering{
       \includegraphics[scale=0.85]{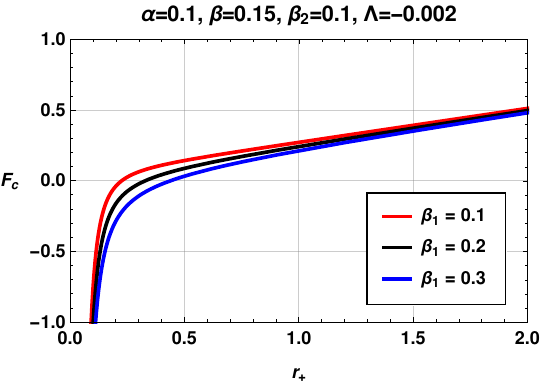} \hspace{5mm}
       \includegraphics[scale=0.85]{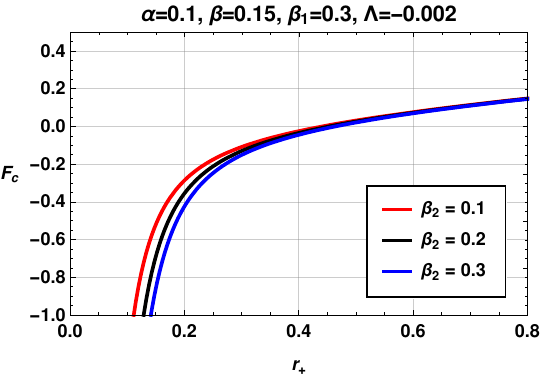}
       
       }
      	\caption{Variation of the corrected Helmholtz energy of the black hole with the event horizon radius $r_+$ for different values of the correction parameters.}
      	\label{figFc02}
      \end{figure*}

By utilising the expression for corrected enthalpy of the black hole, we can obtain the corrected volume which can be obtained by using
\begin{equation}
    V_c = \frac{dH_c}{dP} \Bigg|_{S_c = constant}
\end{equation}
The corrected volume from the above definition is found to be
\begin{equation}
    V_c = \frac{4 \pi  r_+^3}{3}-8 \beta _1 r_++\frac{4 \beta _2}{\pi  r_+}.
\end{equation}
Now, from the expression of corrected volume and corrected entropy, we can obtain the corrected Helmholtz free energy by using the following relation:
\begin{equation}
    F_c = - \int S_c dT_H - \int P dV_c.
\end{equation}
From the above expression, we found
\begin{multline}
    F_c = -(4 \pi ^2)^{-1}\Big[\frac{8 \alpha  \beta _2 \log \left(r_+\right)}{\beta ^5}-\frac{8 \alpha  \beta _2 \log \left(\beta +r_+\right)}{\beta ^5}+\frac{\alpha  \beta _2+\beta ^4 \left(\beta _2 \Lambda +\pi  \beta _1 \log (16 \pi )\right)}{\beta ^4 r_+} +\frac{7 \alpha  \beta _2-3 \pi ^2 \alpha  \beta ^4}{\beta ^4 \left(\beta +r_+\right)} \\ +\frac{\alpha  \left(5 \pi ^2 \beta ^4+3 \beta _2\right)}{\beta ^3 \left(\beta +r_+\right){}^2}+\frac{\alpha  \left(\pi ^2 \beta ^4+\pi  \beta _1 \beta ^2 (2+\log (16)+\log (\pi ))+\beta _2\right)}{\beta  \left(\beta +r_+\right){}^4}-\frac{\alpha  \left(11 \pi ^2 \beta ^4+\pi  \beta _1 \beta ^2 (4+3 \log (16 \pi ))-5 \beta _2\right)}{3 \beta ^2 \left(\beta +r_+\right){}^3} \\ +\frac{\pi  \beta _1 \left(\alpha  r_+^2-\beta ^4-4 \beta ^3 r_+-6 \beta ^2 r_+^2+\Lambda  r_+^2 \left(\beta +r_+\right){}^4-4 \beta  r_+^3-r_+^4\right) \log \left(\frac{\left(\beta ^4-\alpha  r_+^2+4 \beta ^3 r_++6 \beta ^2 r_+^2-\Lambda  r_+^2 \left(\beta +r_+\right){}^4+4 \beta  r_+^3+r_+^4\right){}^2}{\left(\beta +r_+\right){}^8}\right)}{r_+ \left(\beta +r_+\right){}^4} \\ +4 \pi  \beta _1 \Lambda  r_+-\frac{2 \beta _2 \Lambda }{r_+}-\pi  r_+ \left(\beta _1 \Lambda  (4+\log (16)+\log (\pi ))+\pi \right)+\frac{\beta _2}{3 r_+^3} -\pi ^2 \Lambda  r_+^3 \Big],
\end{multline}
which is thermal fluctuation dependent. We have shown the variations of $F_c$ for different model parameters in Fig.s \ref{figFc01} and \ref{figFc02}. These plots illustrate how the corrected Helmholtz energy $ F_c $ of a deformed AdS Schwarzschild black hole varies with the event horizon radius $ r_+ $ for different values of model parameters, specifically the deformation parameter $ \alpha $, control parameter $ \beta $, and correction parameters $ \beta_1 $ and $ \beta_2 $. Physically, the corrected Helmholtz energy provides insight into the thermodynamic stability and energy characteristics of black holes in modified gravity frameworks. As $ r_+ $ increases, $ F_c $ generally rises, indicating that the black hole’s thermodynamic potential becomes more stable with size. The deformation parameter $ \alpha $ and control parameter $ \beta $ significantly impact $ F_c $ at small radii, suggesting that black hole energy corrections are more sensitive to the deformation effects and control influences in the strong gravitational regime. The correction parameters $ \beta_1 $ and $ \beta_2 $ further moderate the growth of $ F_c $, reflecting adjustments to the black hole's entropy and temperature, particularly in smaller black holes. These parameter dependencies highlight the importance of modified gravity effects in black hole thermodynamics and imply that deviations from classical black hole behaviour are most pronounced near the event horizon. Consequently, these corrected energy potentials can help us understand stability conditions, phase transitions, and energy balance in black holes governed by modified theories of gravity.

 The second-order corrected internal energy ($U_{c}$) of the black hole is defined by
\begin{equation}\label{34}
U_{c}=H_{c}-PV_{c}.
\end{equation}
The explicit form for our black hole is given by
\begin{multline}
    U_c = \frac{1}{6} \Big[\frac{12 \alpha  \beta _2 \log \left(\beta +r_+\right)}{\pi ^2 \beta ^5}-\frac{12 \alpha  \beta _2 \log \left(r_+\right)}{\pi ^2 \beta ^5}+\frac{\alpha  \beta ^2+3 r_+^2 \left(\alpha +3 \beta ^2\right)+3 \beta  r_+ \left(\alpha +\beta ^2\right)+9 \beta  r_+^3+3 r_+^4}{\left(\beta +r_+\right){}^3} \\ +\frac{\beta _2 \left(\beta ^7+r_+^2 \left(3 \beta ^5-3 \alpha  \beta ^3\right)+r_+^3 \left(\beta ^4-22 \alpha  \beta ^2\right)-30 \alpha  \beta  r_+^4-12 \alpha  r_+^5+3 \beta ^6 r_+\right)}{\pi ^2 \beta ^4 r_+^3 \left(\beta +r_+\right){}^3}+\frac{\alpha  \beta _1 \left(2 r_+-\beta \right)}{\pi  \left(\beta +r_+\right){}^4}\Big].
\end{multline}

\begin{figure*}[t!]
      	\centering{
       \includegraphics[scale=0.85]{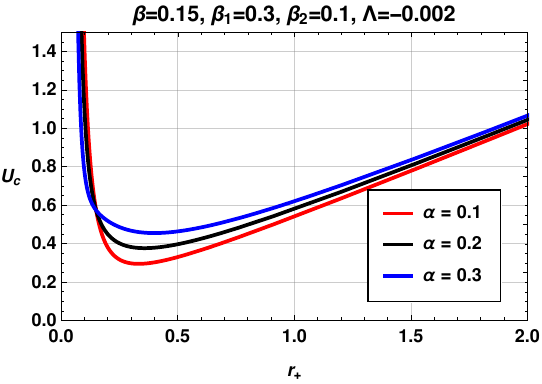} \hspace{5mm}
       \includegraphics[scale=0.85]{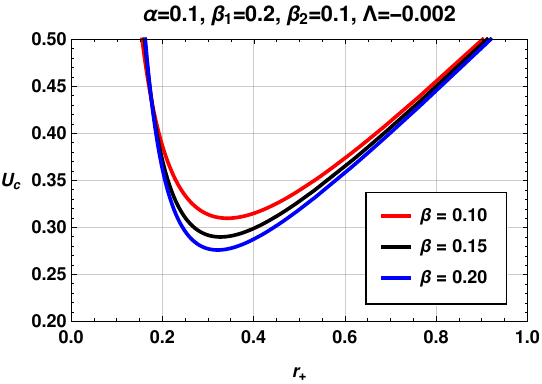}
       
       }
      	\caption{Variation of the corrected internal energy of the black hole with the event horizon radius $r_+$ for different values of the deformation parameter $\alpha$ and control parameter $\beta$.}
      	\label{figUc01}
      \end{figure*}

\begin{figure*}[t!]
      	\centering{
       \includegraphics[scale=0.85]{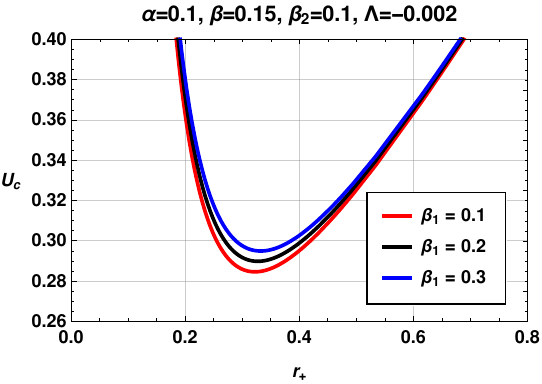} \hspace{5mm}
       \includegraphics[scale=0.85]{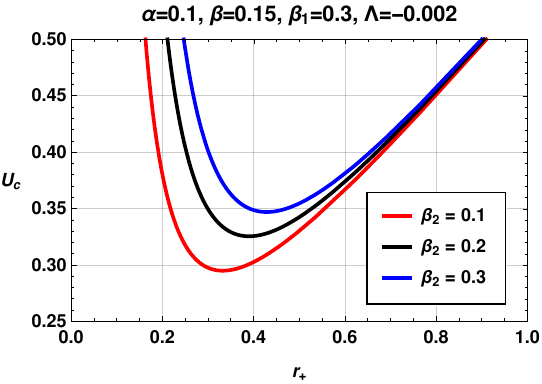}
       
       }
      	\caption{Variation of the corrected internal energy of the black hole with the event horizon radius $r_+$ for different values of the correction parameters.}
      	\label{figUc02}
      \end{figure*}

Fig.s \ref{figUc01} and \ref{figUc02} illustrate the variations of the corrected internal energy $ U_c $ of the deformed AdS Schwarzschild black hole as a function of the event horizon radius $ r_+ $ under different model parameters. In Fig. \ref{figUc01}, we observe the effects of the deformation parameter $ \alpha $ and the control parameter $ \beta $. As shown in the left panel, for fixed $ \beta = 0.15 $, increasing $ \alpha $ results in a slight shift of the minimum internal energy to higher values, suggesting that greater deformation increases the black hole's internal energy. In the right panel, where $ \alpha = 0.1 $, increasing $ \beta $ from $0.1$ to $0.2$ also raises the minimum internal energy and shifts it to a slightly larger horizon radius, indicating that the control parameter influences energy stability at specific radii. Fig. \ref{figUc02} shows the effects of correction parameters $ \beta_1 $ and $ \beta_2 $ on $ U_c $. In both panels, increasing either $ \beta_1 $ or $ \beta_2 $ shifts the minimum internal energy upwards, with a noticeable impact on energy stability around the minimum radius. These trends suggest that both deformation and correction parameters significantly affect the internal energy landscape, which could influence the stability and phase transition behaviour of the black hole. The observed variations in $ U_c $ with different parameters imply that the corrected internal energy is sensitive to modifications in black hole structure, especially in the strong gravity regime near the event horizon.

We shall now explore the impact of thermal fluctuations on Gibbs free energy. In thermodynamics, Gibbs free energy represents the maximum amount of mechanical work that can be extracted from a system. It is mathematically defined by the following relation:
\begin{equation}
G_{c}=M - T S_c.
\end{equation}
By utilizing the corrected values of Helmholtz free energy and volume, we can obtain
\begin{multline}
G_{c}= \frac{1}{12} \Big[\frac{2 \alpha  \left(\beta ^2+3 \beta  r_++3 r_+^2\right)}{\left(\beta +r_+\right){}^3}- 3 x_1^{-1}\Big\lbrace \Big(\beta ^4-\alpha  r_+^2+4 \beta ^3 r_++6 \beta ^2 r_+^2-\Lambda  r_+^2 \left(\beta +r_+\right){}^4+4 \beta  r_+^3+r_+^4\Big) \Big(\beta _1 \log (16 \pi ) \\ -\beta _1 \log \left(\frac{\left(\beta ^4-\alpha  r_+^2+4 \beta ^3 r_++6 \beta ^2 r_+^2-\Lambda  r_+^2 \left(\beta +r_+\right){}^4+4 \beta  r_+^3+r_+^4\right){}^2}{\left(\beta +r_+\right){}^8}\right)+\frac{\beta _2}{\pi  r_+^2}+\pi  r_+^2\Big)\Big\rbrace-2 \Lambda  r_+^3+6 r_+\Big],
\end{multline}
where $x_1 = \pi  r_+ \left(\beta +r_+\right){}^4$

\begin{figure*}[t!]
      	\centering{
       \includegraphics[scale=0.85]{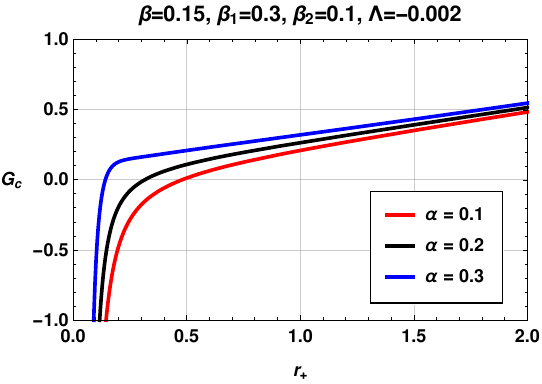} \hspace{5mm}
       \includegraphics[scale=0.85]{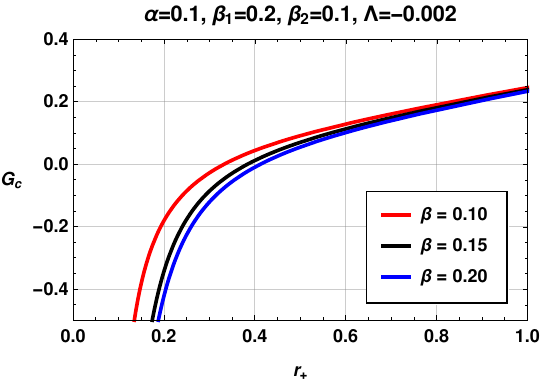}
       
       }
      	\caption{Variation of the corrected Gibbs free energy of the black hole with the event horizon radius $r_+$ for different values of the deformation parameter $\alpha$ and control parameter $\beta$.}
      	\label{figGcr01}
      \end{figure*}

\begin{figure*}[t!]
      	\centering{
       \includegraphics[scale=0.85]{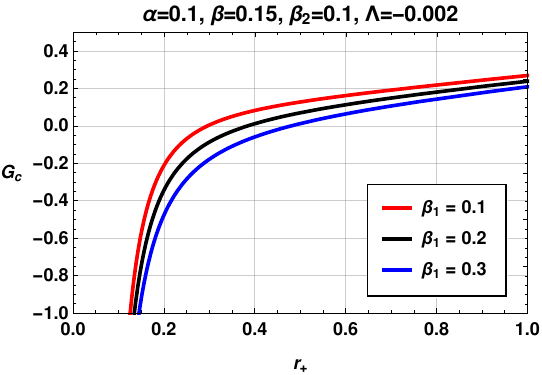} \hspace{5mm}
       \includegraphics[scale=0.85]{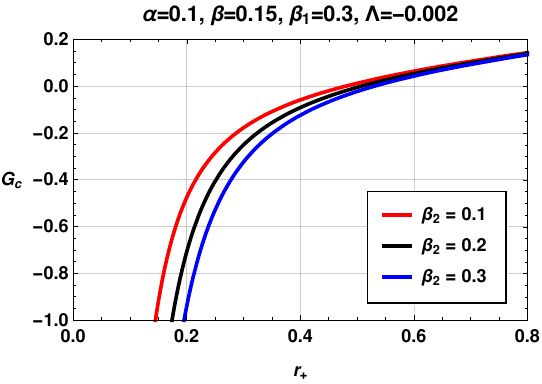}
       
       }
      	\caption{Variation of the corrected Gibbs free energy of the black hole with the event horizon radius $r_+$ for different values of the correction parameters.}
      	\label{figGcr02}
      \end{figure*}

This clearly shows that the thermal fluctuations also affect the Gibbs free energy. To visualise the impacts of the model parameters and the correction parameters, we plot the corrected Gibbs free energy in Fig.s \ref{figGcr01} and \ref{figGcr02}. In Fig. \ref{figGcr01}, one can see that both the deformation parameter $\alpha$ and the control parameter $\beta$ have opposite impacts on the corrected Gibbs free energy $G_c$. An increase in the parameter $\alpha$ increases the corrected Gibbs free energy $G_c$ and the impact is more significant for smaller black holes. On the other hand, an increase in the control parameter $\beta$ decreases the corrected Gibbs free energy $G_c$ and similar to the previous case, the impact is more significant for the smaller black holes. The impacts of the correction parameters $\beta_1$ and $\beta_2$ are shown in Fig. \ref{figGcr02}. One can see that an increase in both parameters decreases the corrected Gibbs free energy $G_c$.
{ An increase in the correction parameters makes the corrected Gibbs free energy $G_c$ more negative, which generally indicates that the black hole is thermodynamically preferred and stable under the given conditions. In other words, a negative $G_c$ suggests that the black hole is not only resistant to decay or evaporation but also represents the most dominant configuration in the system. This stability arises from the fact that a negative Gibbs free energy corresponds to a lower free energy state, making black hole formation thermodynamically advantageous. Consequently, the presence of thermal fluctuations enhances the stability of smaller black hole systems.}

\section{ Phase Transitions and Stability in the presence of Thermal fluctuations}\label{sec6}


Phase transitions in black hole thermodynamics play a crucial role in understanding the stability and evolution of black holes under varying external conditions. These transitions are analogous to those found in conventional thermodynamic systems, where critical phenomena such as the change from one phase to another—solid to liquid, or liquid to gas—are determined by variables like temperature and pressure. In black hole physics, the concept of phase transitions was first explored in the context of the Hawking-Page phase transition, where a black hole in AdS space undergoes a transition between a stable large black hole phase and a radiation-dominated phase. This phase transition occurs at a specific temperature known as the Hawking temperature, which can be influenced by factors such as the black hole's mass, charge, and cosmological constant. Recently, modified gravity theories and quantum corrections have introduced new parameters that affect black hole thermodynamics, leading to more intricate phase transition behaviours. In particular, the introduction of deformation parameters in deformed AdS-Schwarzschild black holes provides a deeper insight into how black holes stabilize or destabilize thermodynamically, as these parameters influence critical temperature thresholds and the overall thermal stability of the system. This section explores these phase transitions in detail, focusing on how Gibbs free energy and its corrected version help identify critical points and transitions in black hole thermodynamics. Moreover, this section also investigates the specific heat of the black hole in the presence of the thermal fluctuations to see how these fluctuations affect the stability and second-order phase transitions of the deformed black hole.

The figures provided illustrate the variation of Gibbs free energy ($G$) and corrected Gibbs free energy ($G_c$) with the Hawking temperature ($T_H$) for deformed AdS-Schwarzschild black holes, highlighting the influence of deformation and thermal correction parameters on the thermodynamic properties of the black hole. These curves provide critical insights into the black hole's phase transitions and stability under varying conditions, particularly focusing on the Hawking-Page phase transition.

\begin{figure*}[t!]
      	\centering{
       \includegraphics[scale=0.85]{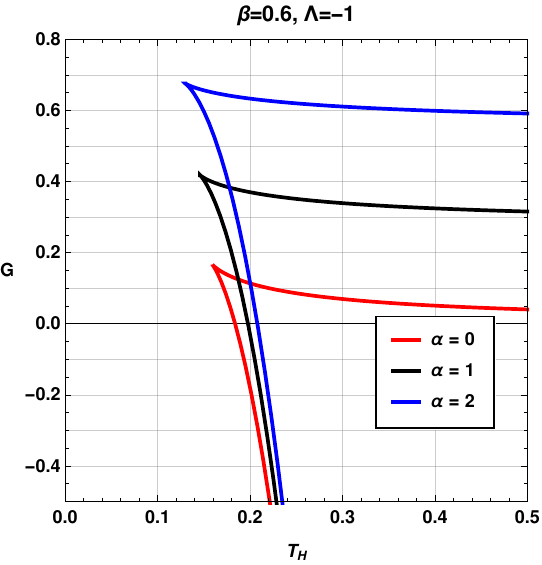} \hspace{5mm}
       \includegraphics[scale=0.85]{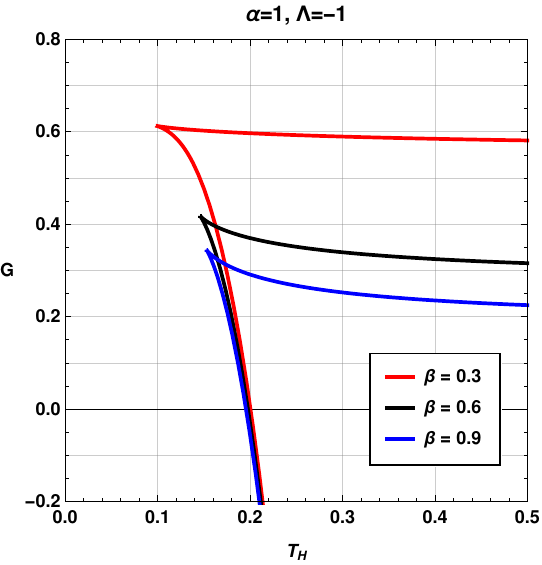}
       
       }
      	\caption{Variation of the Gibbs free energy with the Hawking temperature $T_H$.}
      	\label{figG01}
      \end{figure*}

In Fig. \ref{figG01}, we observe the Gibbs free energy plotted against the Hawking temperature for different values of the deformation parameter $\alpha$ and the control parameter $\beta$, with a cosmological constant $\Lambda = -1$. The left panel explores how the Gibbs free energy changes as $\alpha$ varies, with $\beta$ fixed at $0.6$. As $\alpha$ increases from $0$ to 2, t$h$e curves shift, indicating that higher values of $\alpha$ push the phase transition to occur at higher values of $T_H$. This shift represents the system's increasing thermal stability with higher deformation. The sign change in the Gibbs free energy, where it transitions from positive to negative, marks the Hawking-Page phase transition, signalling the change from a stable black hole phase to an unstable radiation-dominated phase. As the deformation parameter grows, this transition happens at higher temperatures, indicating that larger deformations stabilize the black hole thermodynamically. On the right panel of Fig. \ref{figG01}, $\alpha$ is held constant at $1$ while $\beta$ varies. As $\beta$ increases from $0.3$ to $0.9$, the critical temperature at which the Hawking-Page phase transition occurs shifts to slightly lower values. This suggests that a decrease in $\beta$ also contributes to stabilizing the black hole by broadening the temperature range over which the system remains thermodynamically stable.

\begin{figure*}[t!]
      	\centering{
       \includegraphics[scale=0.85]{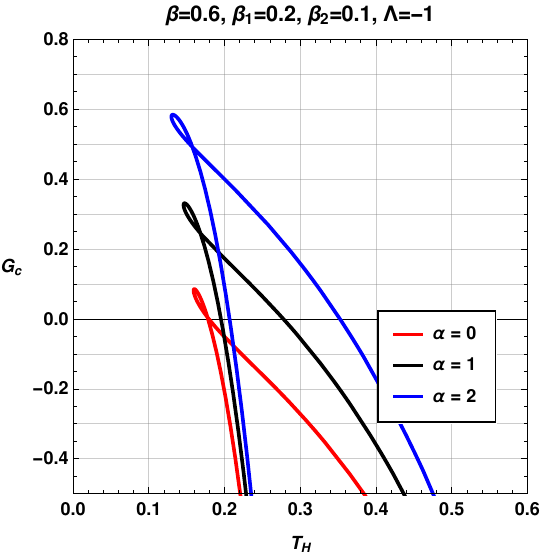} \hspace{5mm}
       \includegraphics[scale=0.85]{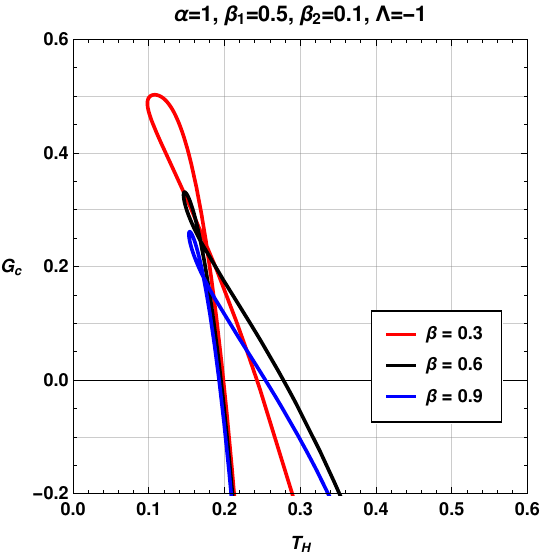}
       
       }
      	\caption{Variation of the corrected Gibbs free energy with the Hawking temperature $T_H$.}
      	\label{figGc01}
      \end{figure*}

In Fig. \ref{figGc01}, the corrected Gibbs free energy ($G_c$) is plotted as a function of the Hawking temperature, considering thermal corrections through the parameters $\beta_1$ and $\beta_2$. The left panel examines the variation of $G_c$ for different values of $\alpha$, with the correction parameters set at $\beta_1 = 0.2$ and $\beta_2 = 0.1$. As the deformation parameter increases, the system stabilizes, with the divergence points shifting toward higher Hawking temperatures. This indicates that thermal fluctuations have a more pronounced effect on smaller black holes, where the corrected Gibbs free energy displays noticeable changes. However, for larger black holes, the effect of thermal corrections is diminished, and the phase transition structure remains largely unchanged. The right panel of Fig. \ref{figGc01} investigates the variation of $G_c$ with $\beta$ for fixed $\alpha = 1$, $\beta_1 = 0.5$, and $\beta_2 = 0.3$. Similar to the previous observations, increasing $\beta$ shifts the divergence points to the right, meaning that the black hole becomes more stable at higher temperatures, with the thermal corrections having minimal impact on the overall phase transition points.

\begin{figure*}[t!]
      	\centering{
       \includegraphics[scale=0.85]{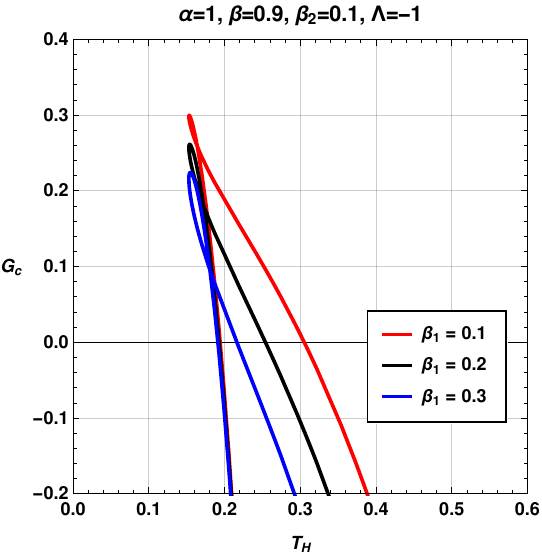} \hspace{5mm}
       \includegraphics[scale=0.85]{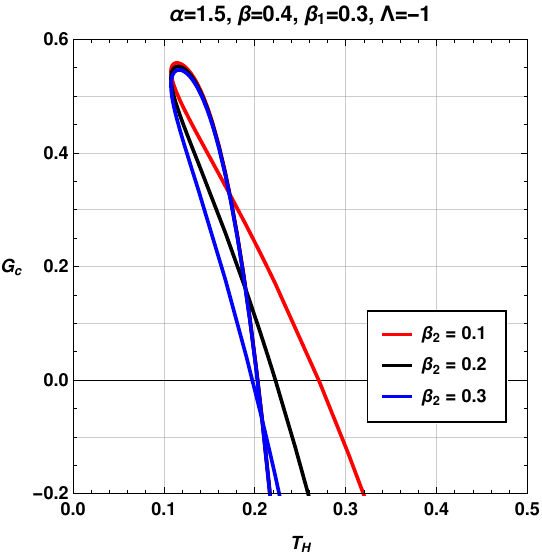}
       
       }
      	\caption{Variation of the Gibbs free energy with the Hawking temperature $T_H$.}
      	\label{figGc02}
      \end{figure*}

Fig. \ref{figGc02} further explores the corrected Gibbs free energy with variations in the thermal correction parameters $\beta_1$ and $\beta_2$. The top left panel demonstrates that for smaller black holes, thermal corrections introduce more noticeable changes in $G_c$, as seen by the spread of the curves for different values of $\beta$. In contrast, larger black holes are less affected by these corrections, and the phase transition remains robust against fluctuations. The top right panel provides similar behaviour, emphasizing that while the thermal correction parameters slightly shift the physical limitation points, they do not alter the fundamental thermodynamic behaviour of the system, particularly the second-order phase transitions. This highlights the resilience of the deformed AdS-Schwarzschild black hole’s thermodynamic properties in the presence of thermal fluctuations and parameter variations.

Overall, these figures illustrate the complex interplay between the deformation parameters ($\alpha$), control parameter $\beta$, and thermal correction parameters ($\beta_1$, $\beta_2$) in governing the stability and phase transitions of deformed AdS-Schwarzschild black holes. The Hawking-Page phase transition, marked by the crossing of Gibbs free energy from positive to negative, serves as a crucial indicator of black hole stability. The stability of the system is enhanced by increasing both the deformation parameter and thermal corrections, indicating strong thermodynamic resilience against phase transitions.

      \begin{figure*}[t!]
      	\centering{
       \includegraphics[scale=0.85]{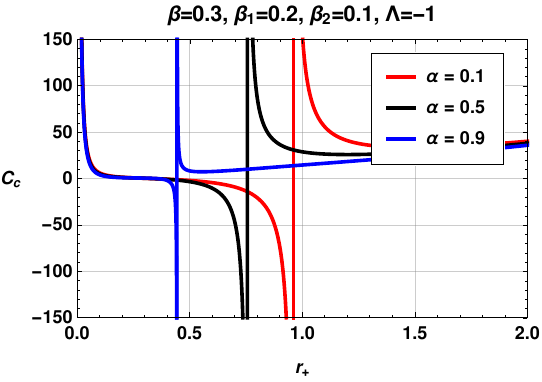} \hspace{5mm}
       \includegraphics[scale=0.85]{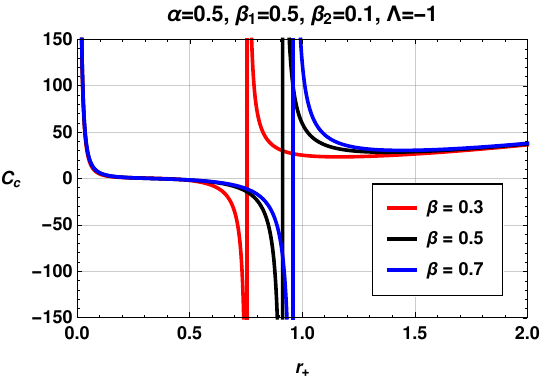}
       
       }
      	\caption{Variation of the corrected specific heat of the black hole with the event horizon radius $r_+$ for different values of the deformation parameter $\alpha$ and control parameter $\beta$.}
      	\label{figCc01}
      \end{figure*}

      \begin{figure*}[t!]
      	\centering{
       \includegraphics[scale=0.85]{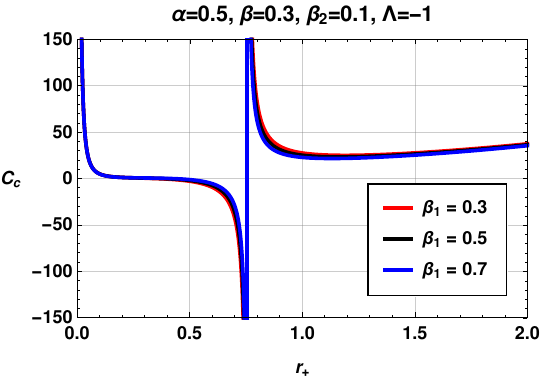} \hspace{5mm}
       \includegraphics[scale=0.85]{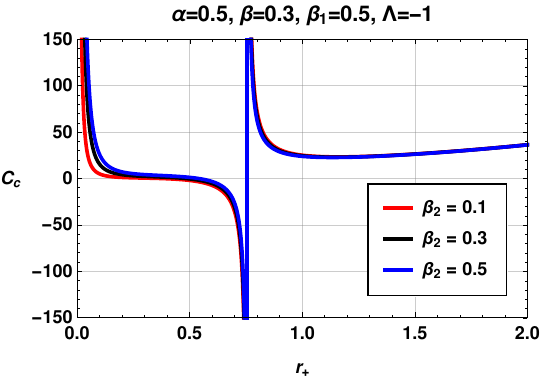}
       
       }
      	\caption{Variation of the corrected specific heat of the black hole with the event horizon radius $r_+$ for different values of the correction parameters.}
      	\label{figCc02}
      \end{figure*}

To study the stability of the black hole system, we investigate the nature of its specific heat. The behaviour of specific heat is a crucial indicator in determining whether the black hole undergoes a phase transition. Specifically, a positive value of the specific heat ensures that the black hole is thermodynamically stable, meaning it can maintain equilibrium and resist phase transitions. In contrast, a negative value of specific heat signifies that the system is unstable, leading to potential phase transitions or even collapse.

In our study, we estimate the specific heat by considering thermal fluctuations in the system. These fluctuations introduce corrections to the classical specific heat, and our goal is to derive an expression that accounts for these effects. Importantly, when thermal fluctuations are negligible (i.e., $\beta_1=\beta_2 = 0$), the corrected specific heat must reduce to the uncorrected classical specific heat, providing consistency with established thermodynamic behaviour.

From the perspective of classical thermodynamics, the specific heat ($C_{c}$) is calculated using the following standard relation:

\begin{equation}\label{Cc}
C_{c}=T_{H}\frac{dS_{c}}{dT_{H}}.
\end{equation}

where $T_{H}$ is the Hawking temperature and $S_{c}$ is the entropy of the black hole. This formula serves as the foundation for our analysis, allowing us to probe the stability and potential phase transitions of the black hole by examining how the specific heat behaves under various conditions.

For this black hole, the corrected specific heat is found to be:
\begin{multline}
    C_c =  2 Q_1^{-1} \Big[\beta _2 \left(\beta +r_+\right) \left(\beta ^4-\alpha  r_+^2+4 \beta ^3 r_++6 \beta ^2 r_+^2-\Lambda  r_+^2 \left(\beta +r_+\right){}^4+4 \beta  r_+^3+r_+^4\right)+\pi ^2 \Lambda  r_+^6 \left(\beta +r_+\right){}^5 \\ -\pi ^2 r_+^4 \left(\beta +r_+\right) \left(\beta ^4-r_+^2 \left(\alpha -6 \beta ^2\right)+4 \beta ^3 r_++4 \beta  r_+^3+r_+^4\right)+2 \pi  \beta _1 r_+^4 \left(\alpha  \left(r_+-\beta \right)-\Lambda  \left(\beta +r_+\right){}^5\right)\Big],
\end{multline}
where $Q_1 = \pi  r_+^2 \left(\beta ^5+\alpha  \beta  r_+^2-3 \alpha  r_+^3+5 \beta ^4 r_++10 \beta ^3 r_+^2+10 \beta ^2 r_+^3+\Lambda  r_+^2 \left(\beta +r_+\right){}^5+5 \beta  r_+^4+r_+^5\right).$

If we set $\beta_1=\beta_2 = 0$
we obtain the uncorrected specific heat of the black hole given by
\begin{equation}
    C_0 = \frac{2 \pi  \Lambda  r_+^4 \left(\beta +r_+\right){}^5-2 \pi  r_+^2 \left(\beta +r_+\right) \left(\beta ^4-r_+^2 \left(\alpha -6 \beta ^2\right)+4 \beta ^3 r_++4 \beta  r_+^3+r_+^4\right)}{\beta ^5+\alpha  \beta  r_+^2-3 \alpha  r_+^3+5 \beta ^4 r_++10 \beta ^3 r_+^2+10 \beta ^2 r_+^3+\Lambda  r_+^2 \left(\beta +r_+\right){}^5+5 \beta  r_+^4+r_+^5}.
\end{equation}

\begin{table}[h!]
    \centering
    \begin{tabular}{|c|c|c|c|c|c|c|c|c|}
        \hline \hline
        $\alpha$ & $\beta$ & $\beta_1$ & $\beta_2$ & $\Lambda$ & $r_{pl1}$ & $r_{pl2}$ & $r_{pl3}$ & $N_{plp}$ \\ \hline \hline

      \hspace{0.3cm}  $0.3$   \hspace{0.3cm}  &   \hspace{0.3cm} $0.5$   \hspace{0.3cm} &   \hspace{0.3cm} $0.0$   \hspace{0.3cm} &   \hspace{0.3cm} $0.1 $  \hspace{0.3cm} &   \hspace{0.3cm}$ -0.002$   \hspace{0.3cm} &   \hspace{0.3cm} $0.317$   \hspace{0.3cm} &   \hspace{0.3cm}   $-$ \hspace{0.3cm} &   \hspace{0.3cm}  $-$   \hspace{0.3cm} &   \hspace{0.3cm} $1$\hspace{0.3cm} \\ \hline
       $ 0.3$  & $0.5$ & $0.3$ & $0.1$ & $-0.002$ & $0.315$  & $-$ & $-$ & $1$ \\ \hline
        $0.3 $ & $0.5$ & $0.6$ & $0.1$ & $-0.002$ & $0.312$  & $-$ & $-$ & $1$ \\ \hline
        $0.3$  & $0.5$ & $0.6$ & $0.0$ & $-0.002$ & $-$   & $-$ & $-$ & $0$ \\ \hline
        $0.3$  &$ 0.5$ & $0.6$ & $0.3$ & $-0.002$ &$ 0.416$  & $-$ & $-$ & $1$ \\ \hline
        $0.3$  & $0.5$ & $0.6$ & $0.6$ & $-0.002$ & $0.497$  & $-$ & $-$ & $1$ \\ \hline
        $0.9$ & $0.0$ & $0.2$ &$ 0.1$ & $-0.002$ & 0.231  & $1.006$ & $-$ & $2$ \\ \hline
       $ 0.9$ & $0.2$ & $0.2$ & $0.1$ &$ -0.002$ & $0.083$  & $0.267$ & $0.572$ & $3$ \\ \hline
        $0.9$ &$ 0.4$ & $0.2$ & $0.1$ & $-0.002$ & $0.311$  &  $-$ & $-$ & $1$ \\ \hline
        $0.0$ & $0.2$ & $0.2$ & $0.1$ & $-0.002$ &$ 0.317 $ & $-$ & $-$ & $1$\\ \hline
       $ 0.4$ & $0.2$ & $0.2$ & $0.1$ & $-0.002$ & $0.350$  & $-$ & $-$ & $1$ \\ \hline
        $0.8$ & $0.2$ & $0.2$ & $0.1$ & $-0.002$ & $0.096$  & $0.256$ &$ 0.520$ & $3$\\ \hline

        $0.1$ & $0.3$ & $0.2$ & $0.1$ & $-1$ &$ 0.328 $ & $-$ & $-$ & $1$\\ \hline
       $ 0.5$ & $0.3$ & $0.2$ & $0.1$ & $-1$ & $0.335$  & $-$ & $-$ & $1$ \\ \hline
        $0.9$ & $0.3$ & $0.2$ & $0.1$ & $-1$ & $0.354$  & $-$ &$-$ & $1$\\ \hline

    \end{tabular}
    \caption{Locations of physical limitation points}
    \label{tab1}
\end{table}

\begin{table}[h!]
    \centering
    \begin{tabular}{|c|c|c|c|c|c|}
        \hline \hline
        $\alpha$ & $\beta$ & $\beta_1$ & $\beta_2$ & $\Lambda$ & $r_{div}$  \\ \hline \hline

      \hspace{0.3cm}  $0.1$   \hspace{0.3cm}  &   \hspace{0.3cm} $0.3$   \hspace{0.3cm} &   \hspace{0.3cm} $0.2$   \hspace{0.3cm} &   \hspace{0.3cm} $0.1 $  \hspace{0.3cm} &   \hspace{0.3cm}$ -1$   \hspace{0.3cm} &   \hspace{0.3cm} $0.962$   \hspace{0.3cm} \\ \hline
       $ 0.5$  & $0.3$ & $0.2$ & $0.1$ & $-1$ & $0.756$  \\ \hline
       $ 0.9$  & $0.3$ & $0.2$ & $0.1$ & $-1$ & $0.444$  \\ \hline

       $ 0.5$  & $0.3$ & $0.5$ & $0.1$ & $-1$ & $0.756$  \\ \hline
       $ 0.5$  & $0.5$ & $0.5$ & $0.1$ & $-1$ & $0.913$  \\ \hline
       $ 0.5$  & $0.7$ & $0.5$ & $0.1$ & $-1$ & $0.959$  \\ \hline

       $ 0.5$  & $0.3$ & $0.3$ & $0.1$ & $-1$ & $0.756$   \\ \hline
       $ 0.5$  & $0.3$ & $0.5$ & $0.1$ & $-1$ & $0.756$   \\ \hline
       $ 0.5$  & $0.3$ & $0.7$ & $0.1$ & $-1$ & $0.756$  \\ \hline

       $ 0.5$  & $0.3$ & $0.5$ & $0.1$ & $-1$ & $0.756$ \\ \hline
       $ 0.5$  & $0.3$ & $0.5$ & $0.3$ & $-1$ & $0.756$   \\ \hline
       $ 0.5$  & $0.3$ & $0.5$ & $0.5$ & $-1$ & $0.756$  \\ \hline

       $ 0.5$  & $0.3$ & $0.5$ & $0.5$ & $-0.002$ & $22.329$  \\ \hline

    \end{tabular}
    \caption{Locations of diverging points}
    \label{tab2}
\end{table}

A critical aspect of analyzing the thermodynamic properties of deformed AdS-Schwarzschild black holes is understanding the behaviour of their heat capacity. The sign of the heat capacity serves as a key indicator of the thermal stability of the system. A positive heat capacity suggests that the black hole is thermally stable, meaning that when energy is added to the system, its temperature increases in a predictable and controlled manner. This indicates that the system can maintain thermal equilibrium. Conversely, a negative heat capacity implies thermal instability, where adding energy to the black hole results in a decrease in temperature, leading to unpredictable and potentially chaotic behaviour. Thus, the sign of the heat capacity is fundamental in determining whether the black hole can resist thermal fluctuations or is prone to instability~\cite{Sahabandu:2005ma, Cai:2003kt}.

Our analysis of the heat capacity, as shown in Fig. \ref{figCc01}, offers significant insights into the black hole's stability by examining the impact of the deformation parameter $\alpha$ and other control parameters. Initially, we observe how varying $\alpha$ affects the heat capacity while keeping other parameters constant. This variation introduces distinct sign-change phenomena in the heat capacity. These changes manifest either through a continuous process at physical limitation points or via discontinuous behaviour at points of divergence. As the parameter $\alpha$ increases, we find that the position of the physical limitation points shifts to larger values, while the divergent points move toward smaller event horizon radii. These behaviours indicate that the black hole remains in a locally stable thermal state, as evidenced by the positive heat capacity.

In contrast, varying the control parameter $\beta$ has a more pronounced effect on the heat capacity, particularly regarding the positions of the physical limitation points. As $\beta$ increases, the divergent points shift toward larger event horizon radii, indicating that the thermal stability of the black hole is influenced significantly by this parameter. This shift highlights the sensitivity of the system to $\beta$ and its role in governing thermal equilibrium.

Furthermore, the introduction of correction parameters $(\beta_1, \beta_2)$ due to thermal fluctuations has minimal but noteworthy effects on the heat capacity (see Fig. \ref{figCc02}). These corrections primarily impact smaller black holes, where the changes in heat capacity are more pronounced. For larger black holes, the effects of $\beta_1$ and $\beta_2$ are less significant. Importantly, the correction parameters slightly modify the positions of the physical limitation points. Specifically, an increase in $\beta_1$ shifts the physical limitation points toward smaller values, while an increase in $\beta_2$ has the opposite effect. Despite these variations, the corrections do not alter the location of phase transitions or critical behaviour within the system. Notably, the second-order phase transition of the black hole remains unaffected by these corrections, reinforcing the robustness of its thermodynamic properties.

The impact of the model parameters on the physical limitation points and diverging points are shown in Tables \ref{tab1} and \ref{tab2}.

In summary, while the deformation and control parameters influence the heat capacity and the system's thermal behaviour, the black hole's fundamental thermodynamic stability remains resilient to these variations. The black hole exhibits consistent stability, even in the presence of thermal fluctuations, further highlighting the robustness of its thermal equilibrium.

\section{Concluding Remarks}\label{sec06}

In conclusion, this paper has provided a comprehensive analysis of the thermodynamic properties and stability of deformed AdS-Schwarzschild black holes. The study primarily focused on the role of deformation parameters ($\alpha$) and thermal correction parameters ($\beta_1$, $\beta_2$) in influencing the black hole's phase transitions and heat capacity. By examining the behaviour of the Gibbs free energy and the corrected Gibbs free energy as functions of the Hawking temperature, we have gained valuable insights into the thermal stability and phase transition characteristics of these deformed black holes.

Our findings demonstrate that the deformation parameter $\alpha$ significantly impacts the Hawking-Page phase transition, with higher values of $\alpha$ leading to an increase in the critical temperature at which the transition occurs. This indicates that the black hole remains thermally stable at higher temperatures when $\alpha$ is larger, suggesting that deformations to the AdS-Schwarzschild black hole enhance its stability. Similarly, the control parameter $\beta$ plays a crucial role in stabilizing the black hole, and its variations lead to shifts in the thermal equilibrium points, as evidenced by the behaviour of the heat capacity.

The introduction of thermal corrections through parameters $\beta_1$ and $\beta_2$ revealed that these corrections have a more pronounced effect on smaller black holes. For larger black holes, the corrections are less significant, and the system remains largely unaffected. However, even for smaller black holes, these corrections do not lead to dramatic changes in the overall thermodynamic behaviour. The corrections slightly modify the physical limitation points, but they do not affect the locations of critical phase transitions or the second-order phase transitions. This robustness underscores the resilience of the black hole's thermodynamic structure, even in the presence of thermal fluctuations.

Moreover, the analysis of the heat capacity further confirms the stability of the black hole system. A positive heat capacity throughout certain ranges of the deformation and correction parameters ensures that the black hole is in a stable thermal state, while any sign change in the heat capacity signals the onset of instability. This careful investigation of the heat capacity, along with the Gibbs free energy analysis, has allowed us to map out the regions of stability and instability for the deformed AdS-Schwarzschild black hole.

In summary, this study has illuminated the critical role of deformation and thermal correction parameters in governing the thermodynamic behaviour of black holes. The ability to enhance black hole stability through these parameters could have significant implications for the study of black hole thermodynamics in various gravitational theories. Future work may focus on extending these results to more complex black hole solutions, such as rotating or higher-dimensional black holes, further advancing our understanding of black hole stability in different theoretical frameworks.

\section*{Acknowledgments}
DJG acknowledges the contribution of the COST Action CA21136  -- ``Addressing observational tensions in cosmology with systematics and fundamental physics (CosmoVerse)".

\section*{Data Availability Statement:}
There are no new data associated with this article.

\section{Conflict-of-Interest Statement:}
The authors declare that they have no conflicts of interest regarding this manuscript.

\section{Funding Statement:}
The authors did not receive any specific funding for this work.


\end{document}